\documentclass[journal=ancac3,manuscript=article]{achemso}

\usepackage[version=4]{mhchem}
\usepackage{graphicx}
\usepackage{amssymb,amsmath}
\usepackage{color}
\usepackage{calc}
\usepackage{xspace}
\usepackage{soul} 
\usepackage [english]{babel}
\usepackage [autostyle, english = american]{csquotes}
\MakeOuterQuote{"}
\usepackage{xr}
\makeatletter
\newcommand*{\addFileDependency}[1]{
  \typeout{(#1)}
  \@addtofilelist{#1}
  \IfFileExists{#1}{}{\typeout{No file #1.}}
}
\makeatother

\newcommand*{\myexternaldocument}[1]{
    \externaldocument{#1}
    \addFileDependency{#1.tex}
    \addFileDependency{#1.aux}
}

\myexternaldocument{supplement}

\author{Amala Raj}
\affiliation[Department of Chemistry]
{Department of Chemistry, Indiana University, Bloomington, IN 47405, U.S.}
\author{William L. Schaich}
\affiliation[Physics Department]
{Physics Department, Indiana University, Bloomington, IN 47405, U.S.}
\author{Bogdan Dragnea}
\affiliation[Department of Chemistry]
{Department of Chemistry, Indiana University, Bloomington, IN 47405, U.S.}

\email{dragnea@iu.edu}
\phone{+1 (812) 8560087}

\title[]
  {Orbital Dynamics at Atmospheric Pressure in a Lensed, Dual-beam, Optical Trap}


\begin{document}


\begin{abstract}
Orbital optical trapping of a dielectric micro-particle in air was studied experimentally using a lensed, counter-propagating dual-beam trap, and by numerical simulations employing ray optics. The essential attributes of particle dynamics are evaluated as functions of the transverse offset between the beams, the axial offset between the laser foci and the total laser power, both experimentally and computationally. We find that the Q-factor of the orbital motion in this previously unexplored scheme is at least two orders of magnitude higher than values attainable with conventional trapping. Under our experimental conditions, silica micro-spheres orbit up to a maximum frequency of $\sim2$~kHz at atmospheric pressure, which can be further increased by increasing the optical power in the trap. With the help of simulations, we discuss how the experimental technique presented here can be further modified to enhance the Q factor of particle's orbital motion. The evolution of orbital frequencies can be a useful signature in analyzing the kinetics of deposition or loss of materials from the surface of levitated particles in a controlled environment. Hence, the approach reported here could find application as an \emph{in situ} single particle technique for probing reactions relevant to atmospheric chemistry.

\end{abstract}


\section{Introduction}

Since its inception in 1970\cite{Ashkin1970}, optical trapping has been widely used as a non-contact and non-invasive method to separate and characterize individual microscopic particles for a broad range of applications\cite{Gao2017,Bradac2018,Bustamante2021}. Optical traps, when coupled with other techniques or applied independently, have proven to be powerful tools in identifying and quantifying the changes in physical parameters associated with a particle interacting with its surrounding medium, for a broad range of phenomena ranging from biomolecular affinity to quantum friction to ultra-precise torque sensors \cite{Block1997,Madsen2021,Ito2007,Gong2018, Gonzalez-Ballestero2021}. Quantitative measurements of torque and other dynamic parameters can be performed with exquisite accuracy if trapping is carried out in vacuum \cite{Stickler2021, Gonzalez-Ballestero2021}. However, because of its relevance for atmospheric processes, there is also great interest in techniques that would allow resolving the state of aerosol particles in interaction with a vapor phase \cite{Krieger:2012iz}. In this work, we isolate and manipulate single particles at atmospheric pressure by employing an orbital optical trap, wherein the particle is moving on a stable, closed trajectory in the field of two counter-propagating, focused laser beams. The stability is due to a balance between optically induced forces
and frictional drag and allows precise measurement of the orbital period, much like a suspension spring in a pendulum clock allows for the accurate measurement of the pendulum period. Through numerical and experimental analyses of particle dynamics in such a trapping scheme, we lay down a formalism to disentangle the key parameters leading to a particle's dynamics. The emerging theoretical framework suggests that such orbital optical traps carry significant promise for real-time, precise measurements of accretion, shrinkage, and oxidative aging of aerosol particles -- phenomena for which discrepancies between various models point to the need for additional experimental data\cite{Houle2021}. 

While control of the rotational motion of a particle resulting from the transfer of spin and/or orbital angular momentum of light have led to a series of interesting microrheological findings \cite{Dholakia2011,Bishop2004, Ahn2020,Asavei2009}, a relatively less explored optomechanical effect is the orbital momentum that can be imparted to a particle by a pair of counter-propagating laser beams, where neither beam carries orbital angular momentum. 
The first experimental demonstration of a dielectric particle moving around a periodic orbit inside a dual-beam trap was acquired with cleaved optical fibers aligned to have an angular offset between them, with water as the medium surrounding the particle\cite{Blakely2008}. Later, this phenomenon was also observed in fiber-optic traps with a transverse offset between parallel beams\cite{Chen2016,Xiao2016,Xu2016}. However, in these experiments, the period was only a few Hz and the quality factor associated with the periodic motion was very low on account of the high viscosity of an aqueous medium. 

Subsequently, orbital motion of dielectric micro-particles was achieved in both air\cite{Li2018} and low vacuum\cite{Zhu2021}, with counter-propagating, dual-beam offset fiber optic traps. While convenient, fiber optic based traps have physical limitations related to geometric constraints, which limit spatial control. Specifically, they are limited to strongly divergent beams only. In addition, the fiber shaft precludes the realization of extended closed orbits, and may induce aerodynamic effects. To address these challenges and explore more widely the possibilities of a free space experiment, we have studied  trapping in air with a lensed, dual-beam assembly employing a pair of focused laser beams for particle control. This new trapping scheme adds another degree of freedom to the toolbox of optical manipulation which allows the particle to move to and around either focus, thus exposing it to both converging and diverging beams. This modification provides an additional mechanism to selectively control the distribution of optical forces in the trap. Moreover, as it will be shown later, a practical advantage of the lensed dual-beam scheme is the access to orbits that are less sensitive to the initial conditions.

We analyze the dynamics of orbital trapped particles by measuring the period, velocities around the trajectory, and the shape of the orbit. Under our experimental conditions, the power spectral distribution of particle's  motion exhibits sharp peaks with narrow widths, at least two orders of magnitude narrower than those associated with periodic oscillation of a trapped particle in air. The apparent quality factors are also considerably higher than what has been reported for  trapping with fiber optic traps in low vacuum\cite{Zhu2021}. Since high mechanical Q factors are desirable for accurate determination of change in the resonance frequency upon particle size and/or mass change\cite{Lavrik2003,Ni2012}, our results have significant implications for the use of optical trapping to probe the kinetics of surface reactions on levitated particles at atmospheric pressure. 

 In addition, with the help of numerical simulations we study the trajectory stability and shape as a function of the location at which particle enters the optical trap, and examine why having control over the point of entry might be beneficial for the implementation of this system as a precision sensor.

\section{Experimental Details}

\begin{figure}[ht]
  \centering
 \includegraphics [width =  \textwidth]{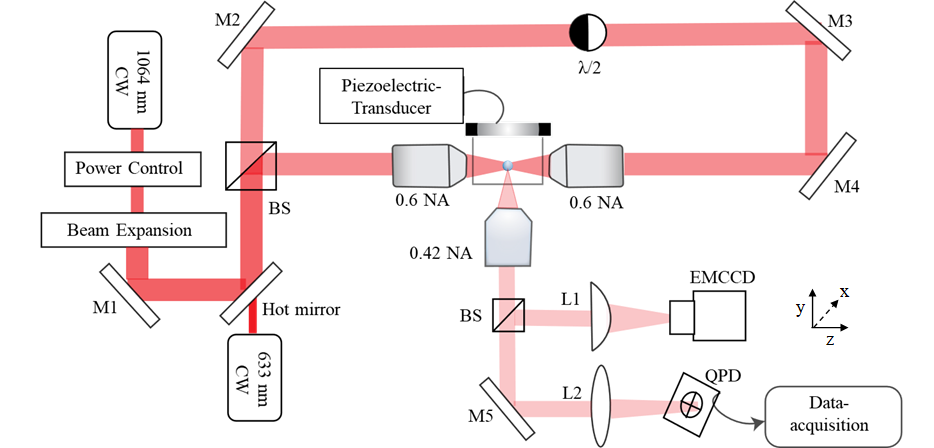}
  \caption{Experimental setup for counter propagating dual-beam trapping in air. Two counter propagating laser beams of equal power and orthogonal polarization states are focused inside a sample chamber where trapping occurs. CW: continuous wave; BS: beam splitter; M: mirror; L: lens; NA: numerical aperture; $\lambda/2$: half-wave plate; QPD: quadrant photodiode; EMCCD: electron multiplier charge coupled device.}
  \label{f:instru}
\end{figure}

A schematic of the experimental apparatus used for dual-beam trapping of particles in air is shown in Figure~\ref{f:instru}. Optical trapping of silica microspheres was performed with a 1064 nm, linearly polarized cw laser. The nearly-Gaussian trapping beam was split into two halves of equal power using a non-polarizing beamsplitter cube and focused by two identical microscope objectives (Nikon 40x Super Plan Fluor ELWD) of working distance 3.5 mm, inside the sample chamber made of glass (wall thickness$\sim$150 $\mu$m). The Gaussian beam waist radius in the focal plane, w$_{0}$, measured by the knife edge method was 1.5 $\mu$m, for both beams. The counter-propagating laser beams were parallel, and their linear polarization states were set to be orthogonal to each other with the help of a half-wave plate, to avoid interference\footnote{In a few cases when the wave-plate was removed, there were no significant differences in the stable orbits.}. The microscope objectives were placed on $xyz$ manual translation mounts with differential micrometers to aid the displacement of beams along the direction of laser propagation ($z$) and in one of the perpendicular directions ($x$). Gravity acts along the -$y$ direction. A low power, visible (632.8 nm) laser for imaging and detection purposes was aligned along the optical axis ($z$).

Silica spheres of diameter 4.82$\pm$0.38 $\mu$m were initially dispersed on a silica glass coverslip. To overcome the attractive van der Waals and capillary interactions between particles and the glass substrate, the coverslip was oscillated with a piezoelectric transducer (70-2221, APC International Ltd.). To achieve the high accelerations necessary to detach the particles from the glass surface, the piezo-transducer was driven with a high power ($\sim 70$ W) square wave at the resonance frequency ($\sim 340$ kHz) of the piezoelectric ring for 1 millisecond. The ultrasonic launcher assembly was placed above the optical trap, Figure~\ref{f:instru}. The coverslip holding the particles and the glass chamber formed a sealed enclosure to ensure protection from air currents. The terminal speed of a descending particle was calculated to be $\sim2$~mm/s.

A long working distance objective (Mitutouyo Plan Apo 20X) was aligned perpendicular to the optical trap axis (along $y$ axis) to collect scattered light from the trapped particle. Half of the scattered light was directed to a quadrant photodiode to measure the displacement of particle along the axis of propagation of laser beams (\(z\)), and along one of the transverse axes (\(x\)). The other half of scattered light was projected onto an EMCCD camera (iXon +897, Andor Technology) to visualize particle motion employing short exposure times and determine the speed of the particle along its trajectory. 

\section{Simulation of Particle Dynamics}

The optical force acting on a particle in a dual-beam optical trap was calculated using a formalism based on the ray (geometric) optics approximation, which is valid when the size of the particle is much larger than the wavelength of illumination. For our case of sphere diameters $\sim5$ $\mu$m, this constraint is marginally satisfied. Under the ray optics approximation, a focused laser beam is considered to consist of a collection of rays and the forces acting on the particle due to individual rays reflecting and/or refracting at surfaces are calculated\cite{Sidick1997,Chen2}. The optical force from a single ray was divided into two orthogonal components which were calculated using the following equations\cite{Sidick1997}.
\begin{equation} \label{eq:dF}
\begin{split}
dF_{s} &=  \hat s\frac{n_1}{c} q_{s} dP\\\\
   dF_{g}  & = \hat g\frac{n_1}{c} q_{g} dP
\end{split}
\end{equation}
where $n_{1}$ is the refractive index of the surrounding medium ($n_{1}=1$ for our case) and \(c\) is the speed of light in vacuum. The unit vectors $\hat s$ and $\hat g$ are  parallel and perpendicular to the ray of light, respectively, and both lie in the plane of incidence. For a spherical particle, the additional rays produced by subsequent reflections and/or refractions all lie also in the same plane of incidence. Since a force on the sphere is determined by a change in ray direction, these forces must also lie in the original plane of incidence and hence can be expanded in terms of $\hat s$ and $\hat g$ alone. The fractions of momentum transferred by light to the particle in the direction parallel and perpendicular to the ray, $q_{s}$ and $q_{g}$, respectively, are given by the following expressions \cite{Ashkin1992}.
\begin{equation} \label{equation3}
\begin{split}
    q_{s} &=  1 + R\cos2\theta_{i} - T^2\frac{cos(2\theta_i - 2\theta_r) + R\cos2\theta_i}{1 + R^2 + 2R\cos2\theta_r}\\\\
    q_{g} &=  -R\sin2\theta_{i} + T^2\frac{sin(2\theta_i - 2\theta_r) + R\sin2\theta_i}{1 + R^2 + 2R\cos2\theta_r}
\end{split}
\end{equation}
where $\theta_i$ and $\theta_r$ are the angle of incidence and angle of refraction of the ray of light. \(R\) and \(T\) are Fresnel coefficients of reflection and transmission, respectively. Rather than sorting out the amount of \(s\) and \(p\) polarization to
associate with each incident ray depending on where it hits the surface, we simply use average values:
\(R=(R_{s} + R_{p})/2\), \(T=1-R\).

In equation~\ref{eq:dF}, \(dP\) denotes the differential power in an incident ray hitting a small patch of surface area (\(dS\)) on the sphere and is given by the following equation:
\begin{equation} \label{equation_4}
    dP=I\cos\theta_i\ dS
\end{equation}
where \(I\) is the intensity of the beam, which we describe with a Gaussian profile.

Although most of the equations we use and present here are from reference~\citenum{Sidick1997}, their version of our equation~\ref{equation_4} is different.  Instead of \(cos(\theta_{i})\), they have the cosine of the sphere's polar angle (measured from the $z$-axis) where it is struck by the incident ray. Since this angle can exceed $90^\circ$, its cosine may be negative which has the unphysical implication $dP<0$. Another difference is that the expression for the parameter $R_{z}$ is attributed to $R_{c}$.

Optical forces from individual rays from both beams were integrated over the whole surface area of the particle illuminated by the beam and their projections along the \(z\) and transverse \(x\) directions were calculated. The total forces when the sphere's center is at \(\vec{r}\) are called axial, \(\hat{z}F_{z}(\vec{r})\),
and transverse, \(\hat{x}F_{x}(\vec{r})\).

A spherical particle moving in an optical trap at the location $\vec{r}$ is also influenced by a viscous drag force from the surrounding medium:
\begin{equation} \label{equation6}
    \vec{F}_{\eta}(\vec{r},t)=-6\pi\eta r_{0} \vec{v}(\vec{r},t)
\end{equation}
where $\eta=1.81\times10^{-5}$Pa$\cdot$s is the dynamic viscosity of air\cite{kaye}, $r_{0}$ is the radius of the silica sphere and \(\vec{v(r,t)}\) is its velocity. The trajectory of a trapped particle was determined by solving Newton's equation of motion in two dimensions.
\begin{equation} \label{eq:Newt}
    m\ddot \vec{r}(t)=\vec{F}(r)+\vec{F}_{\eta}(\vec{r},t) 
\end{equation}
where \(m\) is the mass of the particle.  Others have gone further than our simple treatment.  For instance
one can allow in the medium for random, Langevin forces in addition to viscous drag \cite{Li2018,Zhu2021}. One also can do a full 3D calculation of forces, including forces and motion in the the $y$-direction \cite{Li2018,Zhu2021}. We have done a few 3D calculations but find that the motion in the \(xz\) plane is scarcely changed, with the excursions along $y$ being a small fraction of a micron. The results presented here were all done in 2D, within the \(xz\) plane.

The particle positions at different times separated by small increments ($\delta$t) were computed by numerically solving the finite-difference form of equation \ref{eq:Newt} using a Runge-Kutta $4^{th}$ order approach. Typically, $\delta$t was taken to be 10 $\mathrm{\mu}$s or less. The initial velocity of the particle was kept below 1 $\mu$m/ms in the \(xz\) plane to roughly match its magnitude under our experimental conditions. When the beam powers and focusing lenses are identical, the optical
forces vanish half way between both the two focal planes and the two beam axes. This point is
chosen as the coordinate system origin.  We call $d$ the separation between the axes and $s$ the separation between the focal planes. For $d>0$, the
axis of the beam coming from the right (left) crosses the \(x\)-axis at \(x=d/2 (-d/2)\). We assume the beam on the right (left) has its focal plane at \(z=s/2 (-s/2)\), where $s>0$. These choices suggest that the motion away from the origin should be counterclockwise. This qualitative behavior occurs in all our calculations, after the decay of initial transients. A further consequence of our choices of common beam powers and identical lenses is that the optical forces are antisymmetric under inversion through the origin. This means,
\begin{equation} \label{equationf}
    \vec{F}(x,z)=-\vec{F}(-x,-z)
\end{equation}

The reason for this antisymmetry is that inverting the 2D coordinates of the sphere switches which focal plane and beam axis it is closest to. In all our calculations, $d$ will be positive and, except where explicitly noted, $s$ will be positive too, and usually 35 $\mu m$. The two beam powers are identical, $P_{1}=P=P_{2}$ with $P$ usually 0.25 W. In most of our calculations, the silica sphere's radius ($r_{0}$) is 2.5 $\mu m$

\section{Results and discussion}

\begin{figure}[ht]
  \centering
 \includegraphics [width =  \textwidth]{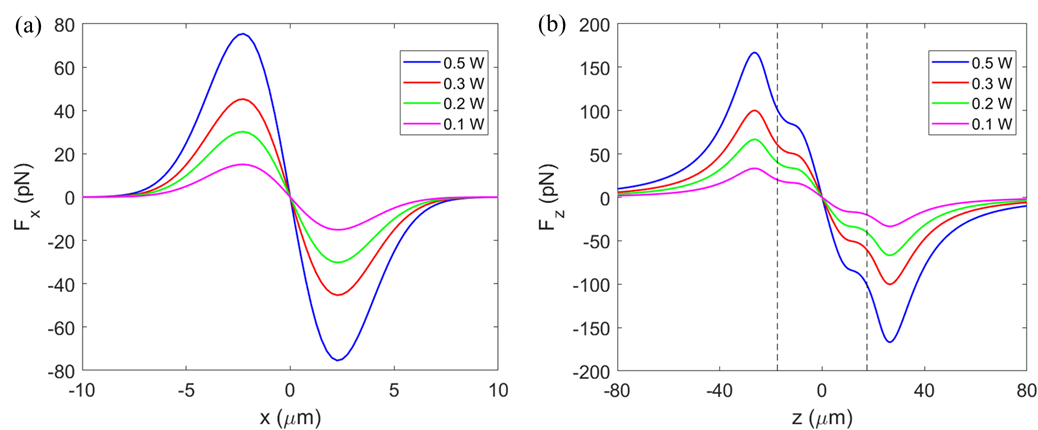}
  \caption{Calculated values of optical forces acting on a silica sphere of radius 2.5 $\mu$m in the transverse (a) and axial (b) directions in the vicinity of the trapping geometry center ((\(x,z\))=(0,0)), at different laser powers, P. The dashed lines indicate the position of the focal planes.}
  \label{f:force}
\end{figure}

The optical forces acting on a silica sphere in a trapping configuration with two collinear ($d$=0, $s$=35$\mu$m) and counter-propagating beams are presented in Figure~\ref{f:force}. As expected, raising the common laser power, $P$, smoothly increases these forces. When the sphere is on the $x$ or $z$ axis, it feels a restoring force only in the transverse or axial direction in Figure~\ref{f:force}a or Figure~\ref{f:force}b. Close to the origin, these forces are linear in the displacement from the origin. Their limiting slopes give an estimate of the trap stiffness which is here $\approx 10^{-5}$~N/m for either direction. If a sphere is released from rest on an axis, it moves along the axis like a damped oscillator, coming to rest at the origin. If the sphere starts instead at an off-axis point, it follows a decaying Lissajous path to the origin. Figures~\ref{f:decay},\ref{f:trap} illustrate these behaviors. Further away from the origin, the magnitude of the transverse force reaches a maximum more quickly than the axial force, which does not have a maximum until $z$ has moved past a focal plane. As the sphere, moving along the $z$-axis and away from the origin, goes through a focal plane, the rays striking it switch from diverging to converging, resulting in additional structure in the axial force profile. Finally, we note that the maximum restoring force along the axial direction is several times higher than the restoring force along the transverse direction, in contrast to single beam optical traps\cite{Rohrbach2005}.

\begin{figure}[ht]
  \centering
 \includegraphics [width =  \textwidth]{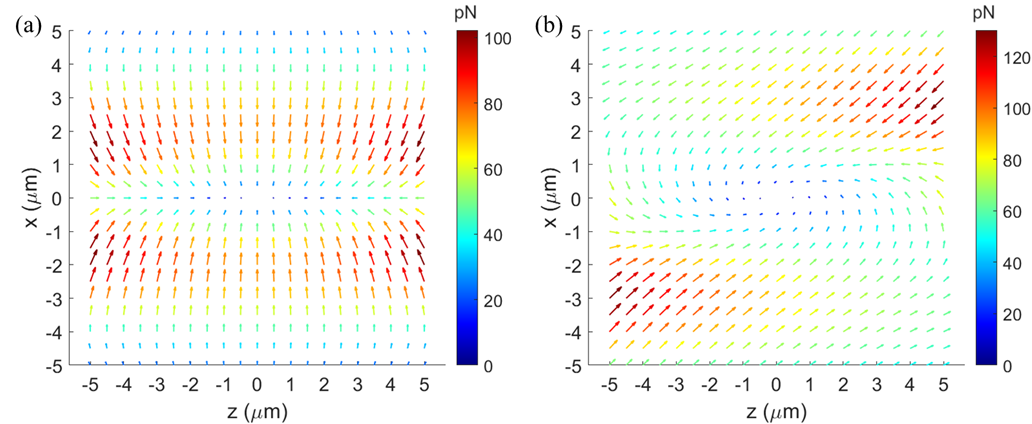}
  \caption{Two-dimensional distribution of optical force field in a dual-beam optical trap with zero transverse offset (a) and for a transverse offset of 3 $\mu$m (b). Axial foci separation, $s$ = +35 $\mu$m, radius of particle = 2.5 $\mu$m. Arrows indicate the direction of force vectors. The optical power from each laser was 0.25 W.}
  \label{f:vector}
\end{figure}

Figure~\ref{f:vector} compares the distribution of optical forces over two dimensions for zero or non-zero beam transverse offset. In both cases, the antisymmetry under inversion of the optical force field is obvious. When the beams are collinear in Figure~\ref{f:vector}a, the optical force is directed mostly towards the center which leads to the motions discussed in the previous paragraph. However when a transverse offset is present as in Figure~\ref{f:vector}b, a “swirling” of the force field becomes evident, suggesting that rotational orbits will be possible. In this case, the curl of the total force is non-zero around the trap center. Thus, the force cannot be generated by a Hamiltonian of the familiar isotropic kinetic energy + scalar potential type \cite{Berry2015}.  A line integral of the optical force on a closed path that surrounds the origin will be non-zero, which means that the forces are non-conservative and that one needs the viscous drag in order to stabilize an orbit. Another subtle feature of Figure~\ref{f:vector}b is that contrary to the impression gained by looking at the red arrows, the beam axes are not at $x$=$\pm3$ $\mu m$  but in fact at $x$=$\pm\frac{d}{2}$=$\pm 1.5$ $\mu m$. Only by going to larger values of $d$ do the separate effects of the two beams become clear. See Figure S3 for the case with $d$=6 $\mu m$ along with the force field profile of a single beam in Figure S4.

\begin{figure}[ht]
  \centering
 \includegraphics [width =1.0 \textwidth]{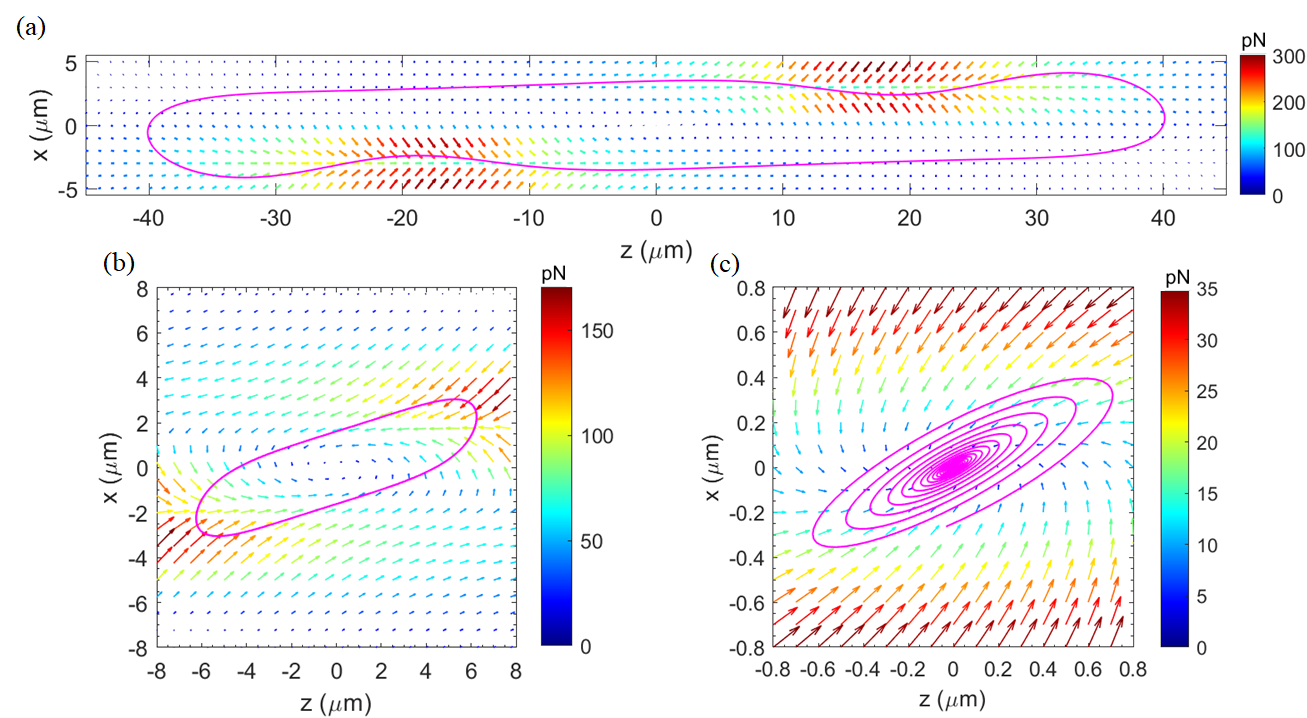}
  \caption{ Simulated trajectories of a silica sphere of radius 2.5 $\mu$m along with the optical force distribution in a counter-propagating dual-beam trap with an axial offset of +35 $\mu$m, and a transverse offset of (a) 6 $\mu$m (b) 4 $\mu$m and (c) 2.3 $\mu$m. The trajectory in c) corresponds to the dynamics of the particle for a duration of 10 ms. The direction of particle's motion is counterclockwise in all three cases. }
  \label{f:trajectory}
\end{figure}

Examples illustrating different stable closed trajectories, after transients (typically less than 5 ms), associated with different transverse offsets are shown in Figure~\ref{f:trajectory}. The common beam power is again 0.25 W. The calculations were all started with the sphere stationary at (\(x,z\))=(0,1 $\mu$m). We have plotted the results in different ways to emphasize different features. For Figure~\ref{f:trajectory}b and~\ref{f:trajectory}c, we have used identical ranges for the $z$ and $x$ axes, adapted to the size of the orbit. This prevents the force vectors from appearing to be rotated due to a stretching of one axis. The force vectors then better appear to be “guiding” the sphere along its trajectory. We have used quotes here because forces only directly produce accelerations, not velocities or locations. In addition, the plotted force field omits the drag force. The orbit in Figure~\ref{f:trajectory}b is confined well inside the focal planes and its frequency is 1587 Hz. The projections of the sphere’s location along the $x$ and $z$ axes were analyzed as a function of time, after it had settled into the stable orbit (Figure~\ref{f:field}b). The trajectory in Figure~\ref{f:trajectory}c is not a stable orbit for our viscosity. However, if the viscosity is reduced by a factor of 2, a stable, nearly elliptical orbit is obtained.

For the case of Figure~\ref{f:trajectory}a, we have stretched the axes to be able to include the whole orbit. The ratio of the orbit’s length/width is nearly 10 and the orbit extends past each focal plane. Such an orbit is physically impossible in a dual fiber trap. There are noticeable dips in the trajectory when the sphere is close to the focal points of the beams. These can be understood by zooming in on sections of the orbit (Figure~\ref{f:zoom}). The orbit’s frequency is 474 Hz. The time-domain signal for the sphere’s location around this orbit is shown in Figure~\ref{f:field}a. The $x$-component shows a fine structure when the sphere is in the region of the dips. The electric field intensity distribution inside the trap corresponding to trapping schemes shown in Figure~\ref{f:trajectory}a,b can be found in Figure~\ref{f:arch}.

There are some common features of the stable, single-loop orbits in Figure~\ref{f:trajectory}a,b. The sphere moves counterclockwise around its orbit and each orbit is centered about the origin. Indeed the orbits have an inversion symmetry about the origin in that if a point $(x,z)$ is on the orbit then so too is the point $(-x,-z)$ and its velocity there is the negative of that at $(x,z)$. These symmetries arise from the antisymmetry of the optical forces, equation~\ref{equationf}, but do not always occur. Orbits that lack such inversion symmetry will be shown later.

\begin{figure}[ht]
  \centering
 \includegraphics [width =  \textwidth]{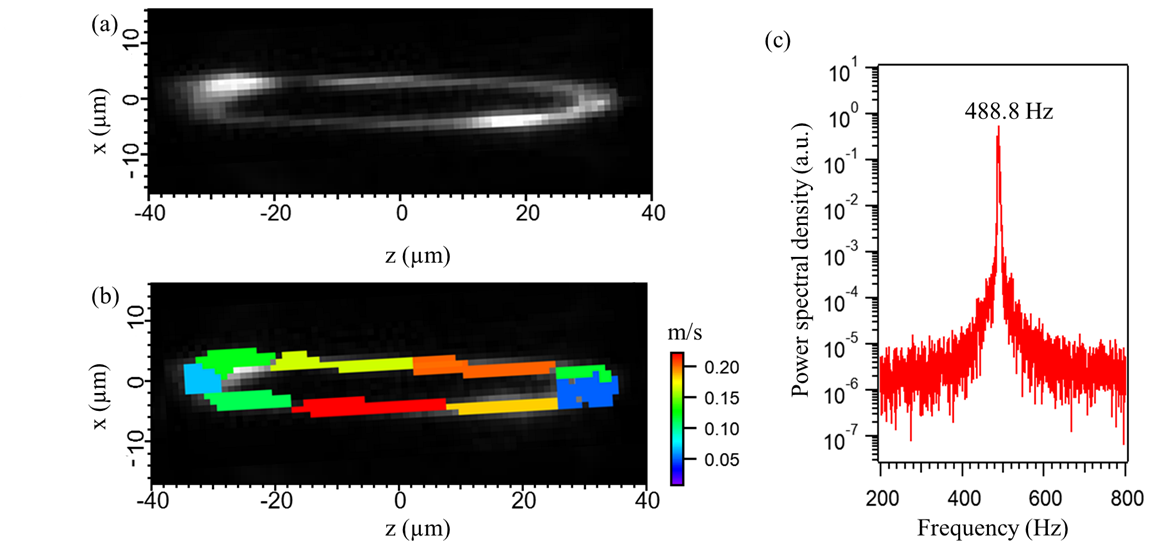}
  \caption{a) EMCCD-acquired experimental trajectory (illustrative of case (a) in Figure~\ref{f:trajectory}). b) Distribution of orbital velocity estimated from short-exposure snapshots. c) Power spectrum derived from particle's motion along the axial direction (\(z)\) by the position-sensitive detector, for the same trajectory.}
  \label{f:data}
\end{figure}

Experimental results for the particle dynamics in a dual-beam trap with a transverse offset of 6 $\mu$m and an axial offset of $+35~\mu$m are presented in Figure~\ref{f:data}. The EMCCD image of particle's orbit acquired at an exposure time of 3 ms/frame is shown in Figure~\ref{f:data}a. The maximum displacements of the particle along axial and transverse axes agree well with the simulation results for the same experimental parameters (Figure~\ref{f:trajectory}a). The speed of the particle at different locations on its trajectory is mapped along the orbit in Figure~\ref{f:data}b. According to our calculations, the particle attains maximum speed in the vicinity of the laser foci, while moving along the dips in the trajectory shown in Figure~\ref{f:trajectory}a. Even though these fine features of the orbit are not well resolved in the EMCCD image, it is evident from the speed distribution that the particle moves the fastest on the quasi-linear segments of the trajectory where the dips are expected. We note that the maximum measured velocity (10's of cm/s), and the particle diameter are similar to values encountered in the updraft of cloud aerosol particles \cite{Guibert2003}, which makes lensed orbital trapping interesting for controlled laboratory experiments of atmospheric relevance.  

The scattered light power spectrum result of the micro-sphere motion along $z$ axis is shown in Figure~\ref{f:data}c. The power spectral distribution was calculated using the following equation:
\begin{equation} \label{equation1}
S_{z}(\omega)={\frac{1}{T_{obs}}}\bigg|\int_{-T_{obs}/2}^{T_{obs}/2} e^{i\omega t}z(t)dt\bigg |^{2}
\end{equation}
where \(S_{z}(\omega)\) is the power spectral distribution corresponding to the sphere's displacement in real time (\(z(t)\)), \(T_{obs}\) is the duration over which the signal is recorded and \(\omega\) is the observation angular frequency. The power spectrum shows a distinct peak corresponding to sphere's orbital frequency at 488.8 Hz, which is consistent with the theoretical prediction.

\begin{figure}[ht]
  \centering
 \includegraphics [width =  \textwidth]{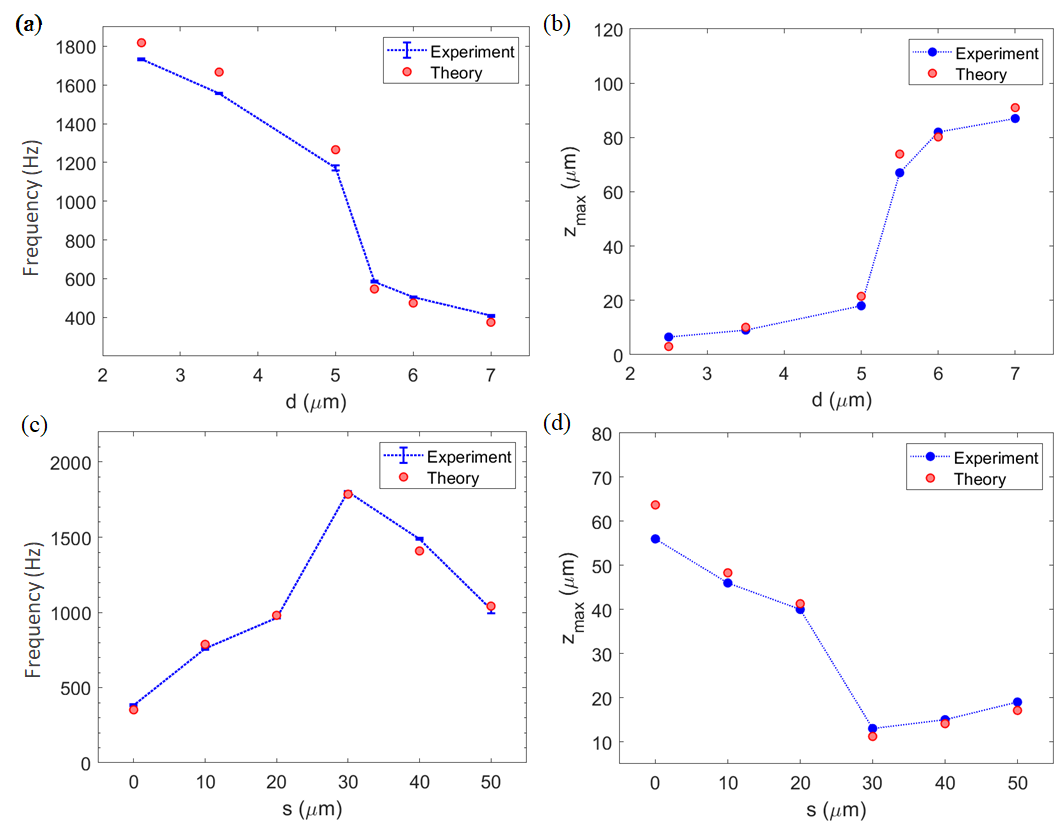}
  \caption{Orbital frequency (a) and maximum axial displacement (b) as a function of the transverse offset between the beams ($s$=35 $\mu$m and \(P\)=0.25 W); Change in orbital frequency (c) and maximum axial displacement (d) upon varying the axial offset between the laser foci ($P$=0.25 W and $d$=3.5 $\mu$m).}
  \label{f:transverse}
\end{figure}

\begin{figure}[ht]
  \centering
 \includegraphics [width =  \textwidth]{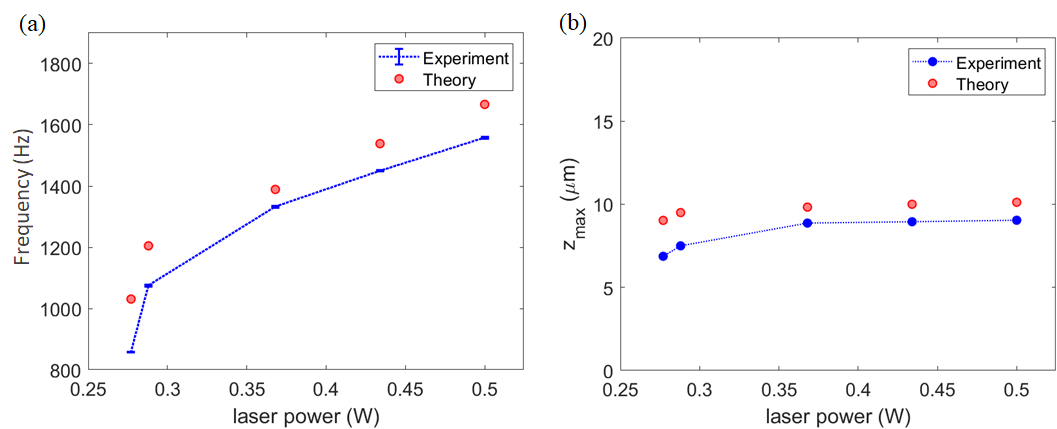}
  \caption{Experimental and numerical analyses of change in orbital frequency (a) and maximum displacement (b) upon changing the total laser power. ($d$=4 $\mu$m and $s$=35 $\mu$m}
  \label{f:axial}
\end{figure}

To further test the computational approach against the experiment, we have examined how the orbital frequency and  spatial extent of the trajectories change separately with transverse offset, axial offset, and laser power, Figure~\ref{f:transverse} and  Figure~\ref{f:axial}. Figure~\ref{f:transverse} a,b shows the changes in orbital frequency and maximum axial displacement of the particle (\(z_{max}=z_{largest}>0 - z_{smallest}<0\)), respectively, upon varying $d$ with $s$ and $P$ held constant.

On increasing $d$, the orbital frequency decreases with a more rapid drop around $d$=5 $\mu$m. In contrast, the orbit size increases with $d$ with a more rapid rise near $d$=5 $\mu$m.
The typical speed of the sphere slightly increases over the range of $d$, so the key to the frequency decrease is the large increase in the orbit size - see Figs.4a,b and below. 

Next, we explored how particle dynamics are changed by variations in $s$ at fixed $d$ and $P$, Figure~\ref{f:transverse}c,d. The changes are not monotonic in $s$: the orbital frequency first increases and then after $s$=30 decreases, while $z_{max}$ has a shallow minimum near $s$=30. Again a rationale is based on considering orbit size. When $s$ is small, $z_{max}$ is considerably larger than $s$ and a large orbit implies (with only a modest increase in speed) that the orbital frequency will be smaller. In the limit of large $s$, the optical forces near the origin become very weak and the viscous drag leads to lower speeds in orbits of nearly constant size, which implies a decrease in orbital frequency. In a lensed, dual-beam trap, both negative and positive axial offsets can be realized. Crossing the two foci for negative axial offsets destabilizes the optical trap leading to particle escape. This result was also predicted by the theoretical model (Figure~\ref{f:escape}).

Lastly, we studied the change in particle response when the common laser power of the beams, $P$, is varied while keeping $d$ and $s$ fixed. Figure~\ref{f:axial}. Upon increasing the optical power, the particle experiences an increased magnitude of optical force resulting in an increase in the orbital frequency. Interestingly, the orbit size quickly reaches a saturation value while the frequency continues to grow. To maintain a balance between optical and drag forces the speeds in an orbit need to increase when $P$ increases, leading to greater frequencies since $z_{max}$ is saturating. With reasonable laser powers, kHz frequencies can be attained inside a dual-beam lensed orbital trap.

To summarize our results so far, there is good agreement between experiments and simulations for the dependencies on $d$, $P$, and $s$, which suggests that the several approximations made in constructing our theoretical model are reasonable.

Since, for a given pressure, the dynamics of the particle in the trap depends on its size, mass, and composition, changes in the dynamics, measurable as frequency shifts and/or orbit shape can be indicative of particular changes in the particle properties. The next question to address is whether orbital trapping could qualify as an analytical technique for the \emph{in-situ} characterization of the chemical and physical changes occurring on single particles in interaction with a gaseous medium. So far, we have seen that one can obtain sharp frequency peaks under experimental conditions that are relevant to atmospheric ones, which have not been explored much, yet (for instance, accretion at drift velocities relevant to updrafts in clouds).

\begin{figure}[ht]
  \centering
 \includegraphics [width =  \textwidth]{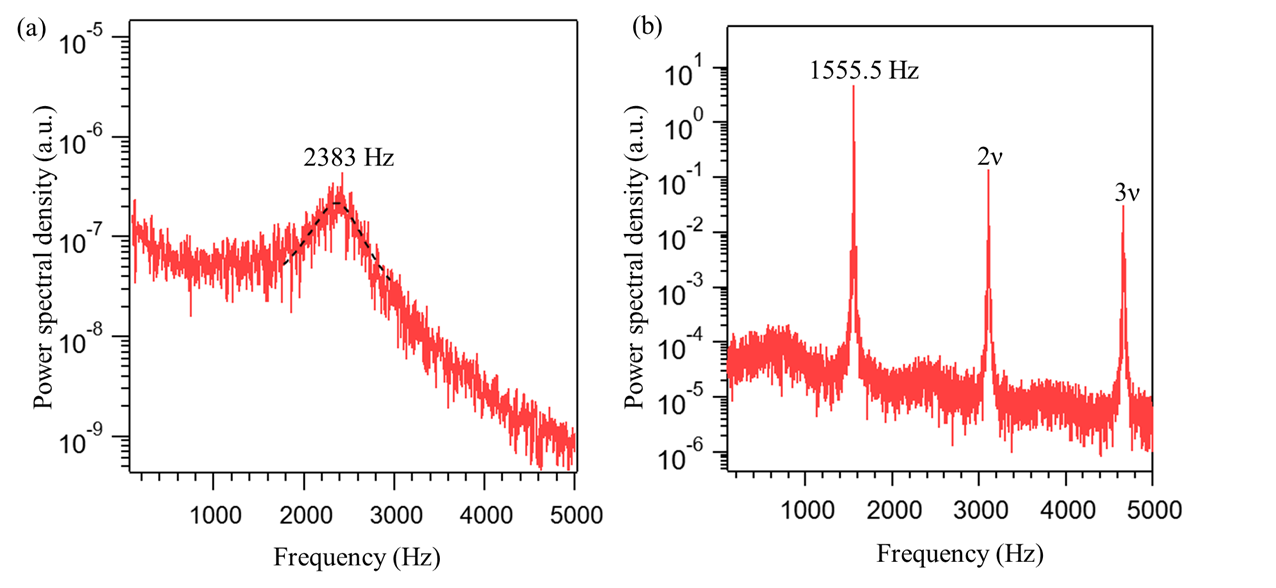}
  \caption{(a) Scattered light power spectral distribution (\(S_{z}(\omega)\)) for the particle's oscillation inside the trap when the laser beams are perfectly collinear. (b) Power spectrum showing the orbital frequency ($\nu$) of trapped particle when the beams have a transverse offset of 3.5 $\mu$m. Peaks corresponding to higher harmonics (2$\nu$, 3$\nu$) of the orbital frequency are also present in the spectrum. For both the cases, $s$= 35 $\mu$m and \(P\)=0.25 W.}
  \label{f:comparison}
\end{figure}

A comparison of power spectra of scattered light collected from a particle, under two different trapping scenarios is given in Figure~\ref{f:comparison}. The power spectrum for periodic oscillation of particle in a dual-beam trap with coaxial beams (Figure~\ref{f:comparison}a) was fit with the following equation\cite{Berg2004,Li2011}\:
\begin{equation} \label{equation8}
S_{z}(\omega)={\frac{2k_{B}T_{0}}{m}}{\frac{\Gamma_{0}}{(\omega^{2}-\Omega^{2})^{2}+\omega^{2}\Gamma_{0}^{2}}}
\end{equation}
The resonance frequency (\(\Omega\)) and the linewidth (\(\Gamma_{0}\)) of the distribution were measured to be 2383 Hz and 530 Hz, respectively. The quality factor for the collinear trap configuration is thus $\sim 4$, Figure~\ref{f:comparison}a. Figure~\ref{f:comparison} b, corresponds to the frequency spectrum of particle dynamics in a dual-beam trap with a transverse offset. The orbital frequency (\(\nu\)) was determined to be 1555.5 Hz, with a linewidth of 10 Hz. The data were acquired over a duration of $2$ s. The quality factor in this case is $\sim 155$. The spectrum also contains peaks for higher harmonics of the fundamental frequency. This is due to the non-harmonic nature of the particle's orbital motion. 

Compared to the oscillatory motion in a traditional trap, the sustained orbital rotation of a particle results in scattered light power spectra with much sharper peaks. Comparable spectral linewidths for conventionally trapped microspheres have only been attained in vacuum conditions\cite{Li2011}.

\begin{figure}[ht]
  \centering
 \includegraphics [width =  \textwidth]{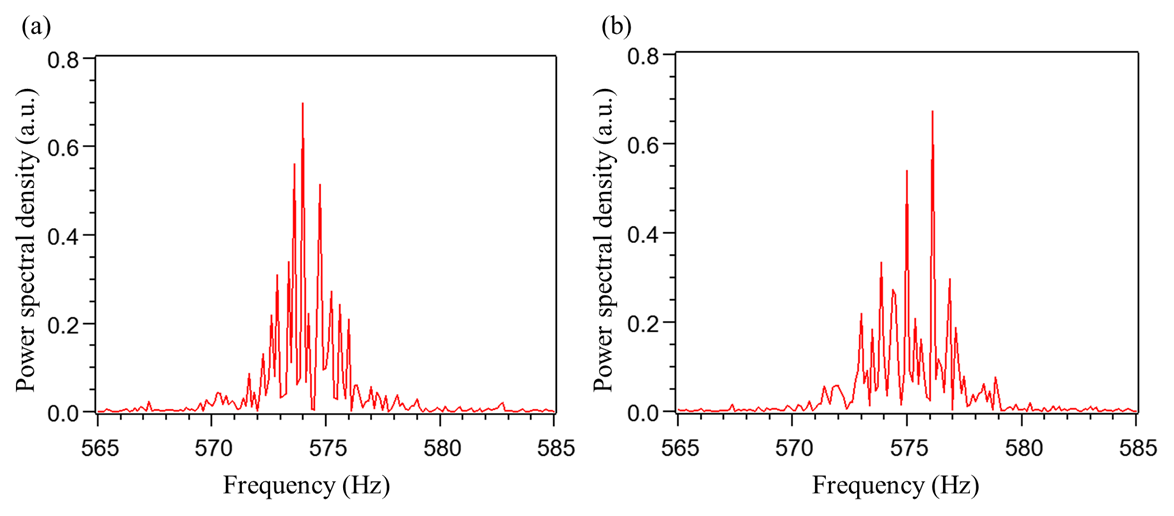}
  \caption{Scattered light power spectrum (\(S_{z}(\omega)\)) for a particle in a dual beam trap with transverse offset ($d$=5.5 $\mu$m, $s$=35 $\mu$m and \(P\)=0.25 W), for the maximum number of data points (total time: 8 s) at maximum sampling rate. (b) Power spectrum corresponding to the dynamics of particle for the next 8 s. A set of narrow peaks can be seen within a frequency range of 10 Hz in both the spectra.}
  \label{f:peaks}
\end{figure}

From the point of view of applications, like for atmospheric aerosol studies, observation of reaction kinetics on levitated particles often require broad time scales, from ms to hundreds of seconds. When using the maximum bandwidth of the acquisition system the observation time was extended to 8 seconds, and the spectral signatures of a particle in an orbital trap with a transverse offset of 5.5 $\mu$m showed an intriguing form of broadening. The spectrum contained a collection of distinct peaks spanning a frequency range of less than 10 Hz, Figure~\ref{f:peaks}. The linewidth associated with the individual peaks from this distribution was about 0.2 Hz. The existence of multiple, well-separated peaks in the power spectrum suggests that there are different "modes" or metastable dynamic states, and the actual trajectory might not be described by a simple single orbit as the calculations suggest. Later on, we will discuss a possible explanation for this result, which suggests a pathway to an even higher quality factor (and thus, sensitivity) for the orbital optical trap. For now, it is worth noting that, if a trapping scheme could be engineered to stabilize the particle in one of the states corresponding to a single narrow peak in Figure~\ref{f:peaks}, it should be possible to obtain very narrow linewidths thus enhancing sensitivity to changes in mass or size of the particle. The scattered light power spectrum under the same experimental parameters, recorded for a wider frequency range can be found in Figure~\ref{f:spectrum}.

Calculations based on the ray optics approximation (Figure~\ref{f:freq}), indicate that the orbital frequency varies nearly linearly with respect to particle radius. For the aforementioned experimental parameters, a change in frequency of 0.2 Hz corresponds to a change in particle radius of 2 nm. However, when the transverse offset between the beams was decreased to 3 $\mu$m, the linewidth associated with the sharpest peak in the distribution (power spectrum given in Figure~\ref{f:peaks_2}) was adequate to resolve a change in radius of 0.5 nm (Figure~\ref{f:freq}b) or about a 0.06\% mass change of particle under consideration. Thus, by tuning the laser parameters, orbital trapping can be made extremely sensitive towards deposition or loss of materials on the surface of a trapped particle.

\begin{figure}[ht]
  \centering
 \includegraphics [width =  \textwidth]{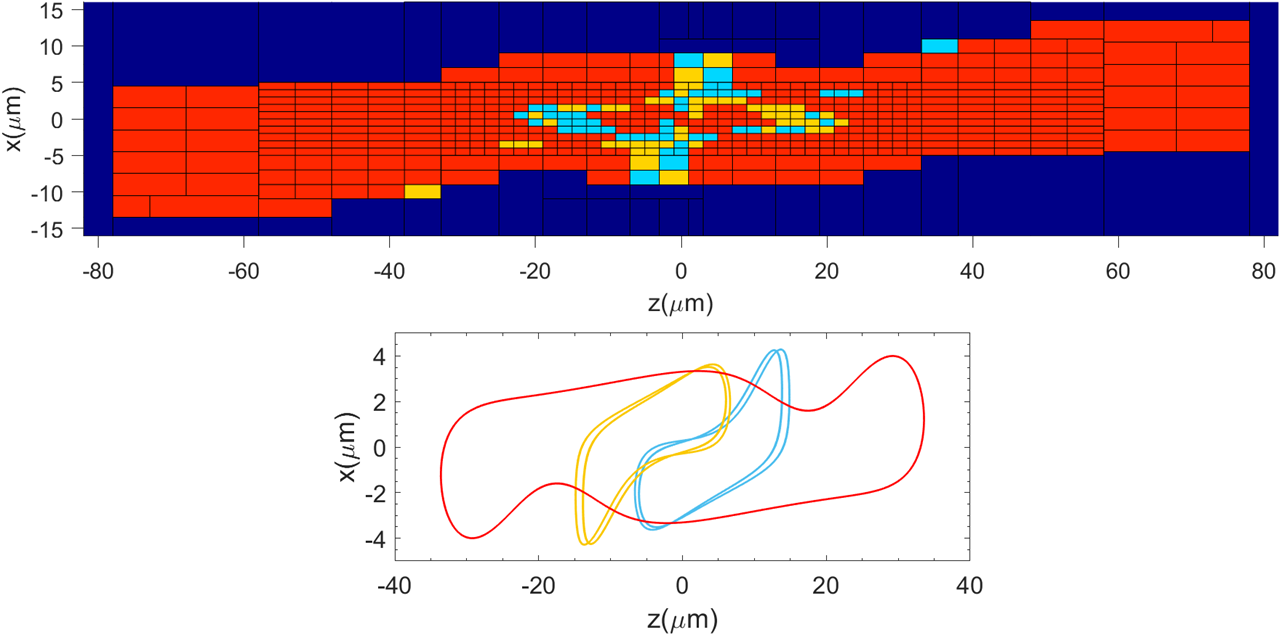}
  \caption{A methodical survey of the dependence on the starting point of a microsphere in the optical field on the trajectory it adopts. Typical orbits along which the particle would advance, if it were to start in the yellow, pale blue or red patches in the block diagram are also displayed in the figure. There are multiple crossovers between the orbits. The regions of instability, where the optical forces are not strong enough to trap the particle, are color-coded in indigo. The yellow and pale blue orbits are asymmetric, and correlated by inversion through the origin. ($d$=5 $\mu$m, $s$=35 $\mu$m and \(P\)=0.25 W)}
  \label{f:mapping}
\end{figure}

To discern the appearance of multiple peaks in the power spectrum, we investigated whether it is feasible for the particle to progress along more than one trajectory under the same experimental conditions. To this end, we simulated the dynamics of a particle by systematically varying its point of entry in the optical field in a dual-beam trap with a transverse offset of 5 $\mu$m, Figure~\ref{f:mapping}. (A similar analysis for a trapping scheme with $d$=4 $\mu$m can be found in Figure~\ref{f:pixelplot}). Our survey results indicate that, depending on where it starts, the particle could be advancing along one of the three distinct periodic orbits (Figure~\ref{f:mapping}) color-coded in red, yellow and pale blue. The regions of instability in the trap are highlighted in indigo. The red orbit has an inversion symmetry and a frequency of 621 Hz. The pale blue and yellow trajectories are degenerate orbits. They both have an asymmetric structure, but are related to each other by inversion through the origin. This means that if a point (\(x,z\)) is on a yellow orbit, the point (\(-x,-z\)) is on a blue orbit. They also differ from the rest of the orbits we have seen so far in that they are double-loop orbits. The particle spends the same amount of time in individual loops of the orbit. The orbital frequency (associated with a single loop) in this case is 1266 Hz. The $z$-projection of these orbits as a function of time are presented in Figure~\ref{f:zpro}. If the particle is initially launched in the area ranging from -30 $\mu$m to +30 $\mu$m in the axial ($z$) direction and -10 $\mu$m to +10 $\mu$m in the transverse ($x$) direction, it will be difficult to predict the exact orbit along which the particle moves during the course of an experiment since even a mere 2 $\mu$m difference in the point of entry along either axis can entirely change the course of the particle. Unfortunately, all three orbits lie in this area, and as evident from Figure~\ref{f:mapping}, there are multiple locations at which the orbits crossover. Hence, even a slight external perturbation, such as thermal fluctuations, could result in the particle migrating randomly from one orbit to another. Thus, we propose that the presence of multiple peaks in the power spectrum can be attributed to orbit cross-over. The real trajectory of the particle may not be a simple, closed orbit as predicted by our theory, which at this time does not include any random, Langevin forces. 

 Our analysis suggests that, in order to obtain superior resolving power, particle motion has to be stabilized along a single orbit. Although there are various sites in the trap where the orbits overlap, the orbital frequency associated with the red trajectory in Figure~\ref{f:mapping} is nearly half of that of the other orbits. Therefore, we hypothesize that, once the particle's dynamics is initiated along the red orbit, shifting to another path might be unfavorable. To attain selectivity towards the red orbit, the particle must be loaded at least 30 $\mu$m away from the origin in the axial direction. With our current experimental setup, preferential launching of particles into specific regions in the trap can not be achieved. Future experiments will address this challenge.
 
 The possibility of more than one type of particle trajectory for the same laser parameters complicates the trends shown in Figure~\ref{f:transverse}. However, the orbits we observe in our experiments always closely match with one of the orbits predicted by the theory in its shape and frequency. Whether other choices of $s$, $d$, $P$ and $r_{0}$ could lead to different types of orbits is yet to be fully explored. A complete optimization of orbital trapping is worth pursuing, in view of the potential of this technology for studying systems mimicking atmospheric aerosol particles in reactive environments. Previously, quantitative measurements of atmospherically relevant reactions on levitated, solid particles were achieved by monitoring the time evolution of Raman signatures of the trapped particles\cite{Rkiouak2014,Tang2014,Gong2019}. Kinetics of evaporation, hydration and coagulation of levitated micro-droplets were studied in general by combining optical trapping with other optical techniques\cite{David2016,Ingram2017,Buajarern2007, Wang2021}. The approach presented here is amenable to in-situ simultaneous spectroscopic and precise physical properties measurements.

\section{Conclusion}

Orbital optical trapping of a micro-particle in air was demonstrated with a lensed, dual-beam trap with a transverse offset. The  characteristics of particle dynamics were analyzed by varying the optical power, the transverse offset between the beams, and the axial offsets between the laser foci, both experimentally and computationally. Under our experimental conditions, the spectral features associated with orbital trapping exhibit higher Q factors than conventional, single focus trapping. Considering that the evolution of orbital frequencies can be correlated to changes in physicochemical properties of the particle in real time, the orbital trapping technique could be applied in the future to probing the kinetics of surface reactions on free, single particles and levitated micro-droplets held in a controlled environment. Once fully optimized, our experimental scheme could provide a powerful, standalone method to explore surface reactions occurring on levitated particles. 

\begin{suppinfo}

Contains details on particle dynamics in collinear dual-beam trap, optical force distribution and electric field intensity mapping for different trapping schemes, scattered light power spectra obtained under different trapping conditions, radius dependence of orbital frequency

\end{suppinfo}

\begin{acknowledgement}
A.R. thanks Dr. Irina Tsvetkova for helpful discussions. The work was partly supported by the U.S. Army Research Office through award \#W911NF1310490.

\end{acknowledgement}

\bibliography{AR_orbitalrotation}

\end{document}


\newpage

\begin{figure}[h!]
  \centering
 \includegraphics [width = 0.6 \textwidth]{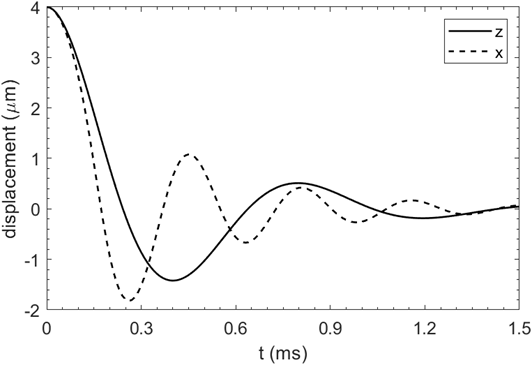}
  \caption{Particle dynamics in a dual-beam trap with two collinear, counter-propagating beams with an  axial offset of +35 $\mu$m between the foci. }
  \label{f:decay}
\end{figure}

\begin{figure}[h!]
  \centering
 \includegraphics [width = 0.6 \textwidth]{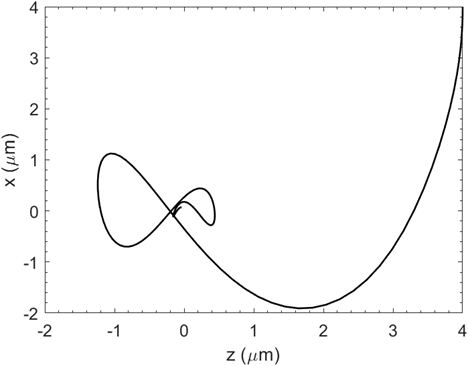}
  \caption{The simulated trajectory of a silica sphere of radius 2.5 $\mu$m in a counter-propagating dual-beam trap with an axial offset of +35 $\mu$m, and a zero transverse offset. }
  \label{f:trap}
\end{figure}
The motions described in the text for the case of two collinear, counter-propagating beams are illustrated in Figures~\ref{f:decay} and \ref{f:trap}. In each, the sphere starts from rest. For Figure~\ref{f:decay}, its starting location is on either the $z$- or $x$-axis and an oscillatory decay to the origin results. For Figure~\ref{f:trap}, the starting location is on a diagonal point with $x=z=4$ $\mu m$. The sphere tries to oscillate at slightly different rates in the $x$ and $z$ directions but they both quickly die out within the 1.5 ms time span shown.

\begin{figure}[h!]
  \centering
 \includegraphics [width = 0.75 \textwidth]{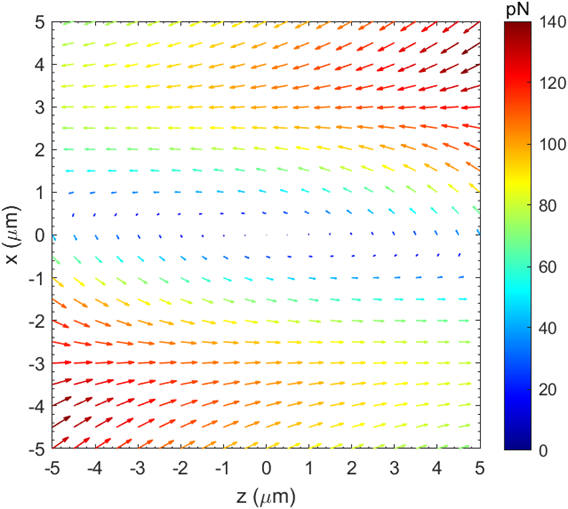}
  \caption{Optical force distribution in a dual-beam trap configuration with an axial offset of +35 $\mu$m, and a transverse offset of 6 $\mu$m.}
  \label{f:dual}
\end{figure}

\begin{figure}[h!]
  \centering
 \includegraphics [width = 0.75 \textwidth]{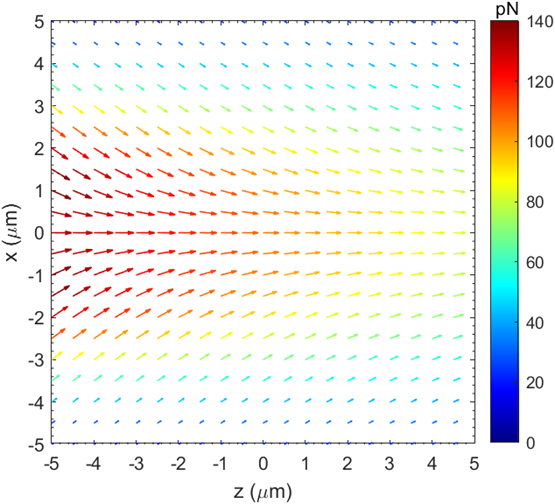}
  \caption{The optical force field generated by a focused laser beam interacting with a silica sphere of radius 2.5 $\mu$m.}
  \label{f:single}
\end{figure}

To supplement the information in Figure~\ref{f:vector}, we show in Figure~\ref{f:dual} how the profiles of the two separate beams become more recognizable when $d$ is increased to 6 $\mu m$ with $s$ still 35 $\mu m$. For comparison, Figure~\ref{f:single} shows, with the same scaling of force vectors, the optical force profile for a single beam coming from the left whose axis is at $x=0$ and whose focal plane is at $z=-17.5$ $\mu m$. A characteristic feature of it is that $F_{x}$ is an odd function of $x$; i.e., the transverse optical force of a single beam is always pulling the sphere towards the beam axis. Coming back to Figure~\ref{f:dual}, note that only for $x$ near zero do the two beams’ forces combine to generate the “swirling” pattern. When $d$ is decreased this “interference” of the two beams extends to larger values of $|x|$, as in Figure~\ref{f:vector}.

\hfill \break

\begin{figure}[h!]
  \centering
 \includegraphics [width =  \textwidth]{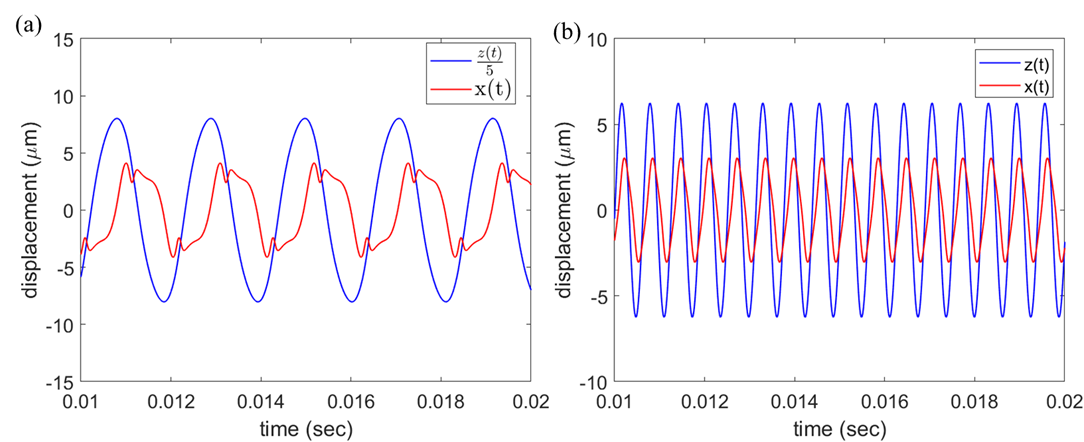}
  \caption{The displacement of particle along $x$ and $z$ axes as a function of time for (a) $d$=6 $\mu$m and (b) $d$=4 $\mu$m. ($s$=35 $\mu$m and \(P\)=0.25 W)}
  \label{f:field}
\end{figure}

The \(x\) and \(z\) projections of the orbits along which the particle moves are illustrated in Figure~\ref{f:field}. The full 2D view of these orbits is in Fig.4a,b. When the transverse offset between the foci is 6 $\mu$m, the maximum displacement along the direction of laser propagation is nearly ten times the maximum displacement along the \(x\) axis, Figure~\ref{f:field}a. The time dependence of the latter is non-harmonic due to the complex shape of the orbit. The jiggles arise from the two dip regions. For a transverse offset of 4 $\mu$m (Figure~\ref{f:field}b), the axial and transverse components of orbits are nearly sinusoidal since the stable trajectory of the particle is closer to an ellipse.

\begin{figure}[h!]
  \centering
 \includegraphics [width =  \textwidth]{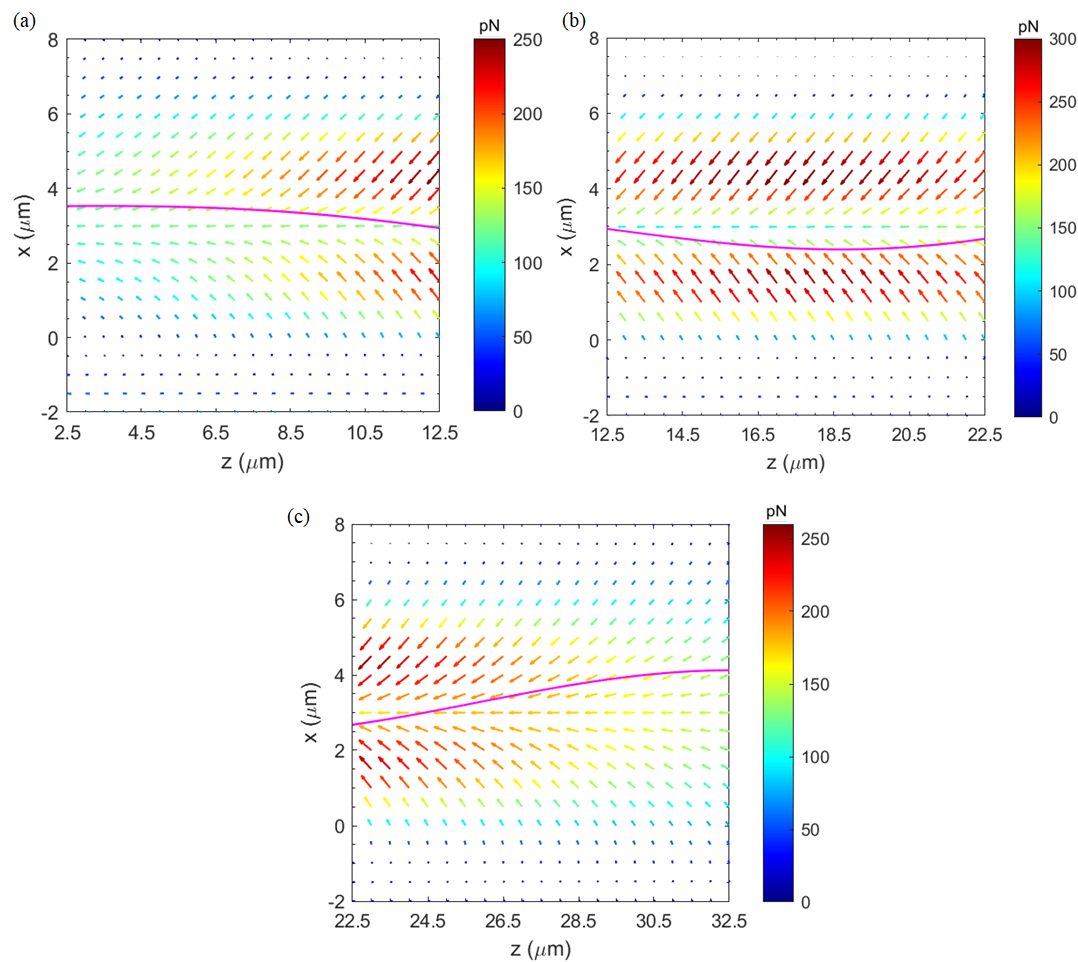}
  \caption{Simulation results showing segments of particle's orbit close to the focus of one of the beams along with the optical force distribution in the vicinity of a laser focus in a dual-beam trap. ($s$=35 $\mu$m, $d$=6 $\mu$m and $P$=0.25 W)}
  \label{f:zoom}
\end{figure}

\hfill \break

In Figure~\ref{f:zoom}, we show three separate zoomed-in plots of a portion of the orbit of Figure~\ref{f:trajectory}a in the region of a dip. As in Figure~\ref{f:trajectory}b, the optical force vectors appear to be “guiding” the sphere along its trajectory. To understand the apparent oscillations of the sphere around the beam axis at $x=3$ $\mu m$, consider separately the optical force components as functions of $z$ along the orbit. $F_{z}$ is always negative and has a local extremum near the focal plane of the beam coming from the right. In contrast, $F_{x}$ oscillates around this beam’s axis (recall the description of Figure~\ref{f:single} above). Using these rough characterizations of the optical forces and ignoring the drag force, one would expect to see an oscillation in $x$ whose period in $z$ is smoothly varying. This behavior is in qualitative agreement with the trajectory segments in Figure~\ref{f:zoom}.
\hfill \break
\hfill \break
\hfill \break
\hfill \break
\hfill \break
\hfill \break
\hfill \break

\begin{figure}[h!]
  \centering
 \includegraphics [width =  \textwidth]{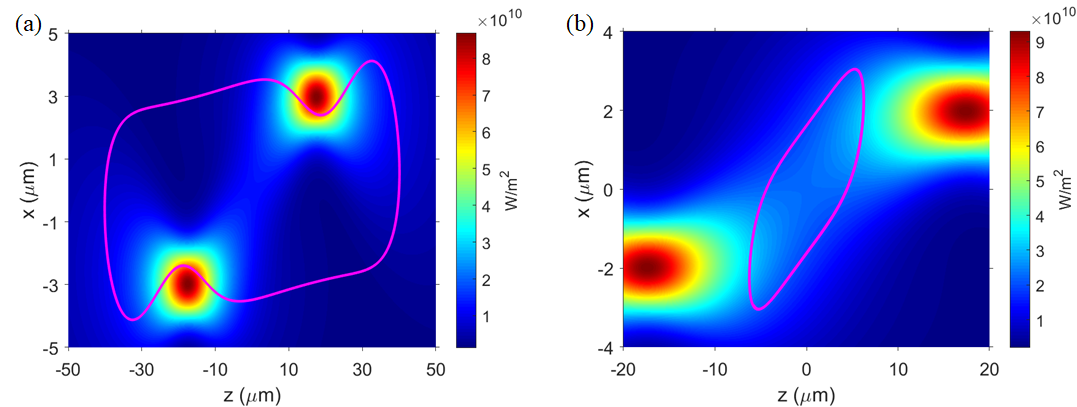}
  \caption{Simulated trajectories of a silica sphere of radius 2.5 $\mu$m along with the electric field intensity map in a counter-propagating dual-beam trap with an axial offset of +35 $\mu$m, and a transverse offset of (a) 6 $\mu$m and (b) 4 $\mu$m.}
  \label{f:arch}
\end{figure}

As an alternative to Figure~\ref{f:trajectory}, we show in Figure~\ref{f:arch} the trajectories of a particle (after transients) under two different trapping configurations along with the distribution of the electric field intensity, $I$, in the trap. The focal points of the two beams are easy to identify but note how the different stretching of the coordinate axes leads to apparent changes in the shape of the regions of high intensity. In the limit when the particle diameter is much less than the wavelength, the optical force on a small particle is attributed to two distinct mechanisms\cite{Harada1996}: 1) a scattering force in the axial direction and proportional to $I$ plus 2) a gradient force proportional to the gradient of $I$. Although our system has large particles compared to the wavelength, it is remarkable how the force profiles we find are qualitatively similar to those obtained in the Rayleigh limit. For instance, the vanishing of $F_{x}$ on the beam axis has an analog in the vanishing of the $x$-derivative of $I$ on the beam axis, while the smooth $z$ variation of $F_{z}$ for a single beam with a peak in the focal plane has an analog in the variation of $I$. However, the similarity between the two limits does not extend to quantitative values. The only remnants of the Rayleigh theory in our analysis are in the labels $s$ and $g$ in equation~\ref{eq:dF} and \ref{equation3}.

\begin{figure}[h!]
  \centering
 \includegraphics [width =   \textwidth]{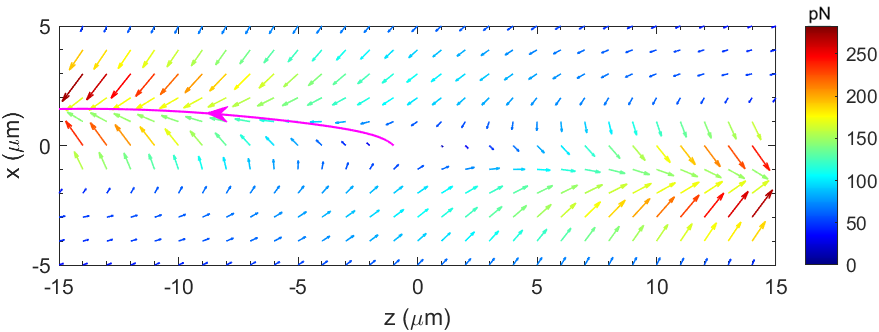}
  \caption{The simulated trajectory of the particle under an experimental scheme in which the laser foci crossover ($s<0$). The calculated optical force distribution in the trap is illustrated in the background. The initial coordinate of the particle is $(x_{1},z_{1})=(0,-1 \mu m)$. ($s$=-35 $\mu$m, $d$=3 $\mu$m and \(P\)=0.25 W)}
  \label{f:escape}
\end{figure}

Figure~\ref{f:escape} illustrates the two-dimensional distribution of optical force fields in a dual-
beam trap when the axial offset between the laser foci is $-35$ $\mu m$. The transverse offset
between the beams is 3 $\mu m$. Upon crossing the foci of the two beams, the optical force vectors point in a direction away from the trap center. A silica micro-sphere of radius 2.5 $\mu m$ is initially launched at the coordinate, $(x,z)$=(0,-1 $\mu m$), in the trap. The particle gets pushed
away from the trap center and eventually escapes.

\begin{figure}[h!]
  \centering
 \includegraphics [width = 0.7 \textwidth]{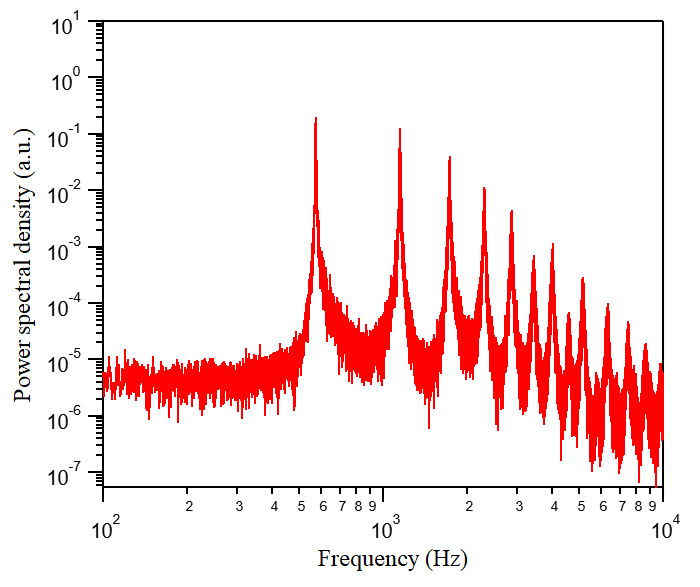}
  \caption{The scattered light power spectrum (\(S_{z}(\omega)\)) of particle trajectory when the transverse offset between the beams was kept at 5.5 $\mu m$. Higher harmonics of the orbital frequency are also present in the spectrum.} 
  \label{f:spectrum}
\end{figure}

\hfill \break
\hfill \break

The scattered light power spectrum(\(S_{z}(\omega)\)) of the particle dynamics (observed for 8 seconds) inside an orbital trap is shown in Figure~\ref{f:spectrum}. The transverse offset between the beams was maintained at 5.5 $\mu m$. The particle orbits at a frequency of $\sim$575 Hz. The spectrum also shows peaks for higher harmonics of the orbital frequency due to non-harmonic nature of the particle trajectory. A spectrum for a narrow range of frequencies around the first harmonic is in Fig.9

\begin{figure}[h!]
  \centering
 \includegraphics [width =  \textwidth]{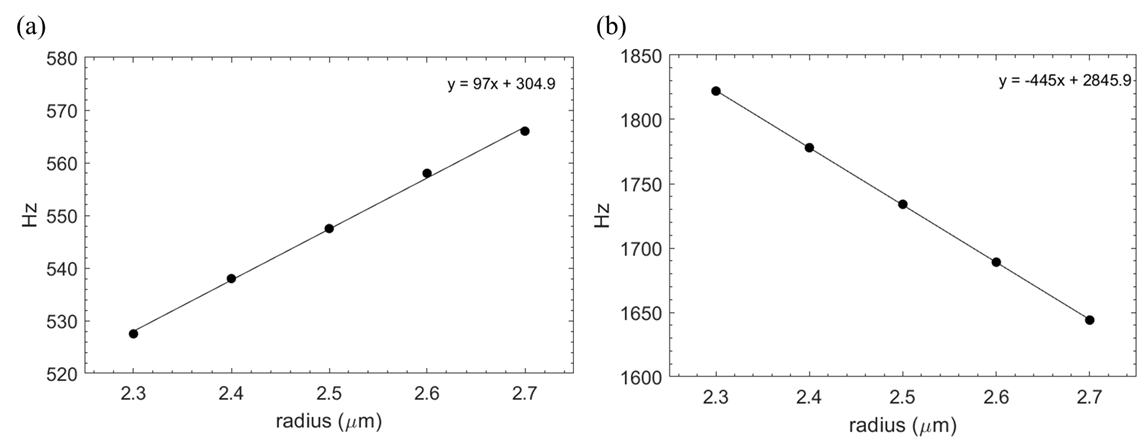}
  \caption{  Radius dependence of orbital frequency for (a) $d$=5.5 $\mu$m and (b) $d$=3 $\mu$m. ($s$=35 $\mu$m and \(P\)=0.25 W)}
  \label{f:freq}
\end{figure}

Figure~\ref{f:freq} shows the calculated values of orbital frequencies as a function of $r_{0}$ for two different choices of $d$. To compute the orbital frequency in each case, the particle’s dynamics was simulated for a duration of 1 second and the $z$ projection of the orbit was fourier transformed. In all the simulations, the particle was started from rest at $(x,z) = (0,1\mu m)$.

Figure~\ref{f:freq}a,b show a remarkable difference in the results which is a challenge to explain. The derivative of the frequency with respect to $d$ is of the opposite sign and nearly five times larger in Figure~\ref{f:freq}b compared to Figure~\ref{f:freq}a. The simple "why" answer is because the orbits for $d=5.5$ $\mu m$ ($3$ $\mu m$) are similar in size and shape to those in Figure~\ref{f:trajectory}a (\ref{f:trajectory}b), respectively. Hence, in one case the trajectories go around each focal plane while in the other they stay in between the focal planes. However, a deeper understanding is not obvious. Changing $r_{0}$ changes the optical force field in a way that is not just a scale factor. Then over the particle's new stable orbit, the net work done by the optical and viscous forces will still balance, but how these two forces now vary around the orbit's slightly different size and shape is not clear, except when looking at their time dependence in the simulation. In addition, changing $r_{0}$ changes the particle's mass, which also modifies the particle's acceleration, velocity and orbit shape. Still, the practical conclusion is that by tuning the transverse offset between the beams, orbital trapping can be made more sensitive to changes in particle size or mass.

\begin{figure}[h!]
  \centering
 \includegraphics[width=0.75
  \textwidth]{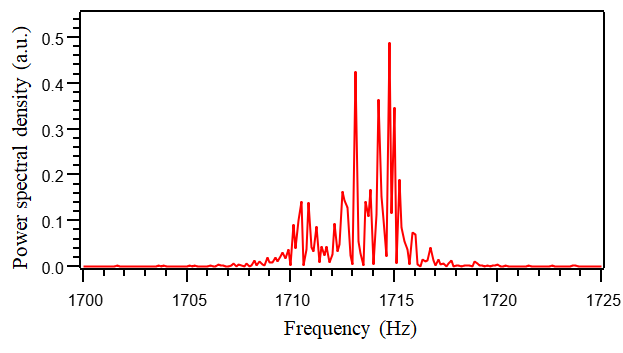}
  \caption{ Power spectral distribution of scattered light from a trapped particle(\(S_{z}(\omega)\)) in a dual-beam trap with a transverse offset of 3 $\mu$m. Within a frequency range of 5 Hz, a group of narrow peaks can be seen in the spectrum. ($s$=35 $\mu$m and \(P\)=0.25 W)}
  \label{f:peaks_2}
\end{figure}

Figure~\ref{f:peaks_2} demonstrates the power spectrum corresponding to particle dynamics in a dual-beam trap with a transverse offset of 3 $\mu$m and an axial offset of 35 $\mu$m. A collection of narrow peaks can be seen within a frequency range of 5 Hz. The sharpest peak in the distribution has a linewidth of about 0.2 Hz. 

\begin{figure}[h!]
  \centering
 \includegraphics [width =  \textwidth]{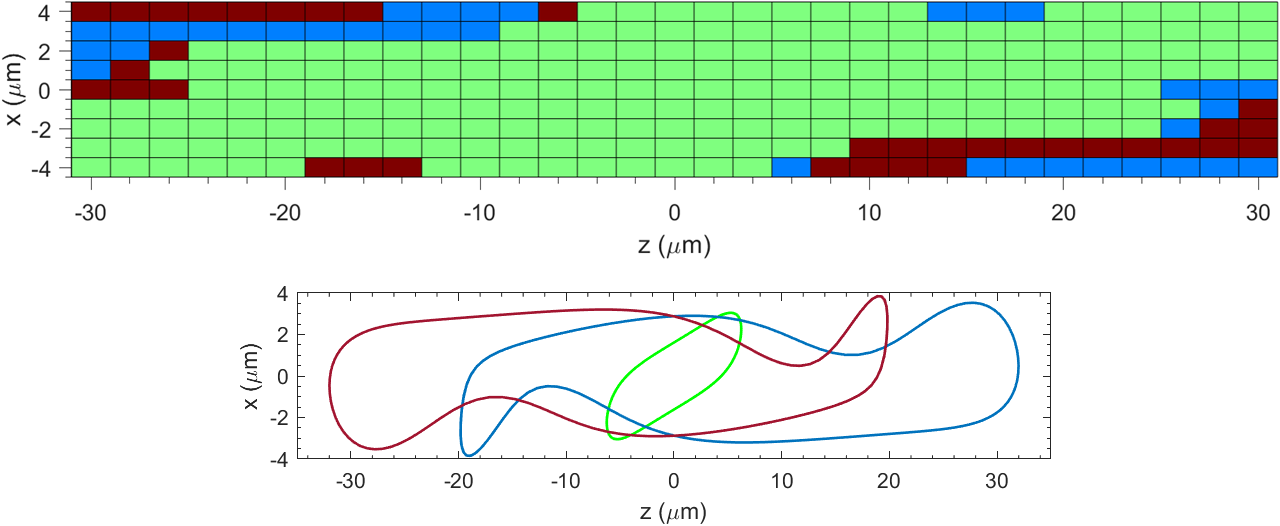}
  \caption{ A map of different possible particle trajectories, in an orbital optical trap with a transverse offset of 4 $\mu$m, as a function of particle's initial coordinate in the trap. Three distinct orbits are possible for this trapping scheme. The blue and maroon orbits are inverses of each other. ($s$=35 $\mu$m, \(P\)=0.25 W)  }
  \label{f:pixelplot}
\end{figure}

Figure~\ref{f:pixelplot} is a survey of possible particle trajectories as a function of particle's point of entry in a dual-beam trap with a transverse offset of 4 $\mu$m. The orbits corresponding to the green, blue and maroon cells in the block diagram are also displayed in Figure~\ref{f:pixelplot}. The green orbit is nearly elliptical and has an inversion symmetry. Its orbital frequency is 1587 Hz. If the particle is launched at a location close to the trap center, there is a high probability of particle advancing along the a green orbit. The maroon and blue orbits are asymmetric but are related to each other by inversion through the origin. Their common  orbital frequency is 749 Hz. Unlike the orbits corresponding to a trapping geometry with $d$=5$\mu m$ (Figure~\ref{f:mapping}), all the orbits shown here are single-loop trajectories. Also, the degenerate orbits in this case have more spatial extent and lower orbital frequency than the symmetric orbit.

\begin{figure}[h!]
  \centering
 \includegraphics [width =  \textwidth]{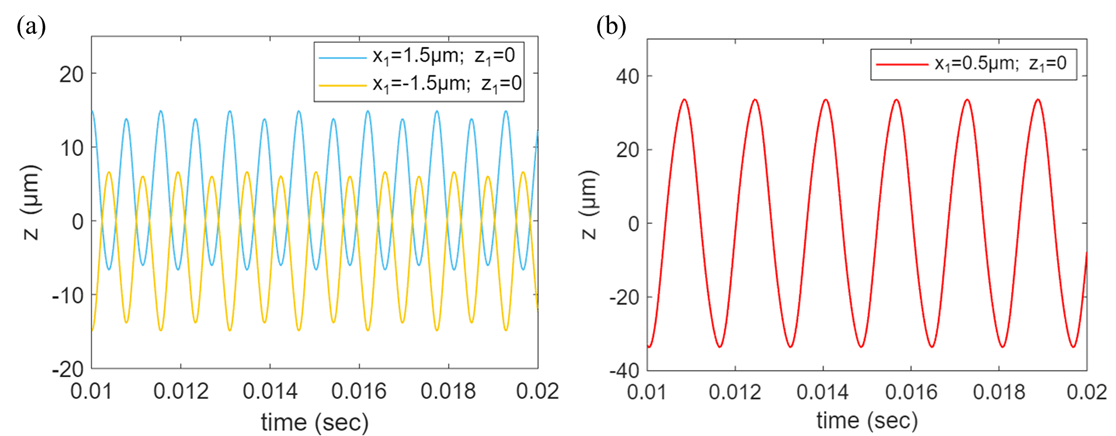}
  \caption{ Projection of particle's orbits onto $z$ axis, as a function of time. The initial coordinates, ($x_{1}$,$z_{1}$) are given in the legend. The initial coordinates for the pale blue and yellow trajectories are inverses of each other. The period of the red trajectory is nearly twice that of blue and yellow ones. ($s$=35 $\mu$m, $d$=5 $\mu$m and \(P\)=0.25 W)  }
  \label{f:zpro}
\end{figure}
The projection of particle's orbits onto \(z\) axis for three distinct trajectories possible inside an orbital trap (transverse offset of 5 $\mu$m and an axial offset of 35 $\mu$m, Figure~\ref{f:mapping}) is shown in Figure~\ref{f:zpro}. Pale blue and yellow trajectories are inverses of each other and are double loop orbits. Their orbital frequencies are equal, Figure~\ref{f:zpro}a. The time period associated with individual orbits of the double-loop trajectory are the same. The red orbit is symmetric under inversion through trap center and has more spatial extent than the other two orbits, Figure~\ref{f:zpro}b. The orbital frequency associated with the red trajectory is close to half that of the pale blue and yellow orbits.

\bibliography{AR_orbitalrotation}